\documentclass[conference,10pt]{IEEEtran}

\input epsf
\usepackage{caption}
\usepackage{graphicx,epstopdf}   
\usepackage{pifont,dsfont}
\usepackage{bm,bbm}
\usepackage{subcaption}
\usepackage{float}
\usepackage[utf8x]{inputenc}
\usepackage{color}
\usepackage{algorithmic,mathtools}
\usepackage[table]{xcolor}
\definecolor{Gray}{gray}{0.9}

\usepackage[linesnumbered,algoruled,boxed,lined]{algorithm2e} 
\usepackage{amssymb,amsmath,amsfonts,amsthm}
\usepackage{cite,citesort}
\usepackage{balance}
\usepackage[utf8x]{inputenc}
\usepackage{graphicx}
\usepackage{epsfig}

\usepackage{mathdots}
\usepackage{mathtools}
\DeclarePairedDelimiter{\ceil}{\lceil}{\rceil}
\DeclarePairedDelimiter{\floor}{\lfloor}{\rfloor}

\usepackage[
top    = 1.70cm,
bottom = 1.05in,
left   = 0.61 in,
right  = 0.61 in]{geometry}

\IEEEoverridecommandlockouts

\newtheorem{theorem}{Theorem}

\newtheorem{corollary}{Corollary}

\newtheorem{ppro}{Proposition}

\begin{document}

\title{Recursive Periodicity Shifting for Semi-Persistent Scheduling of Time-Sensitive Communication in 5G
\vspace*{-0.2cm}
}

\author{
\IEEEauthorblockN{
Nan Jiang, Adnan Aijaz, and Yichao Jin
}
\text{Bristol Research and Innovation Laboratory, Toshiba Europe Ltd., Bristol, United Kingdom}\\
\{nan.jiang, adnan.aijaz, yichao.jin\}@toshiba-bril.com
\\


\vspace*{-0.6cm}
}



\maketitle

\begin{abstract}
Various legacy and emerging industrial control applications create the requirement of periodic and time-sensitive communication (TSC) for 5G/6G networks. State-of-the-art semi-persistent scheduling (SPS) techniques fall short of meeting the requirements of this type of critical traffic due to periodicity misalignment between assignments and arriving packets that lead to significant waiting delays. To tackle this challenge, we develop a novel \emph{recursive periodicity shifting} (RPS)-SPS scheme that provides an optimal scheduling policy by recursively aligning the period of assignments until the timing mismatch is minimized. RPS can be realized in 5G wireless networks with minimal modifications to the scheduling framework. Performance evaluation shows the effectiveness of the proposed scheme in terms of minimizing misalignment delay with arbitrary traffic periodicity. 
\end{abstract}
\vspace*{-0.2cm}
\begin{IEEEkeywords}
5G, 6G, automation, deterministic, uRLLC, SPS, periodic shifting, time-sensitive communication, TSN.
\vspace*{-0.2cm}
\end{IEEEkeywords}

\section{Introduction}
5G wireless networks have been designed with native support for ultra-reliable low-latency communication (uRLLC) which is crucial for supporting critical services across a number of industry verticals. A key requirement for both legacy and emerging control-centric industrial applications is periodic delivery of small-sized data under ultra-low latency and strict timing constraints, which is also referred to as time-sensitive communication (TSC) \cite{aijaz2020private,mannweiler2019reliable}. Realizing TSC at scale and with tight time synchronization (i.e., extreme uRLLC) is also the vision of future 6G systems \cite{park2020extreme}. 

Historically, semi-persistent scheduling (SPS) techniques have been developed to support voice services over cellular networks \cite{jiang2007principle}. SPS provides periodic access to channel resources for deterministic traffic. However, adopting SPS scheme for control-centric TSC applications creates a new challenge wherein the incoming traffic may require extremely high scheduling precision from the time domain perspective which could be beyond the limit of state-of-the-art networks. A key example is integration of 5G and time-sensitive networking (TSN) systems where the external TSN traffic could be arriving in time windows as short as \(10\mu\)s \cite{mannweiler2019reliable}, while the minimum resource unit on the 5G air-interface is on the order of \(100\mu\)s (e.g., 2 symbols with 30 kHz sub-carrier spacing (SCS) covering \(71\mu\)s) \cite{sachs20185g}. Consequently, the period of assignments may not exactly match the period of the incoming traffic, which can lead to increase of the misalignment delay and/or the wastage of channel resource. 



As argued in this work, one may allow a changeable scheduling period varied according to a pre-defined principle that completely solves the period misalignment problem. To do so, a recursive periodicity shifting (RPS) scheme is proposed for SPS in this work, which recursively aligns the scheduling period of resource units until it finds the optimal policy, i.e., each assignment located right after the arrival of the corresponding packet. The proposed scheme is proved to minimize the period misalignment; thus, it can achieve theoretically minimal misalignment delay. The proposed solution can be realized in 3GPP-based systems with minimal modifications.

The rest of the paper is organized as follows. Section II briefly introduces SPS and provides literature review. Section III describes the system model and problem formulation. Section IV elaborates the concept of the RPS scheme. Section V proposes methods to apply the RPS scheme in cellular systems. Section VI presents numerical results, and section VII concludes the paper.

\section{Semi-Persistent Scheduling, Periodicity Misalignment Issue, and Related Prior Works}

\begin{figure}
    \begin{center}
    \begin{minipage}[t]{0.47\textwidth}
    \centering
        \includegraphics[width=1\textwidth]{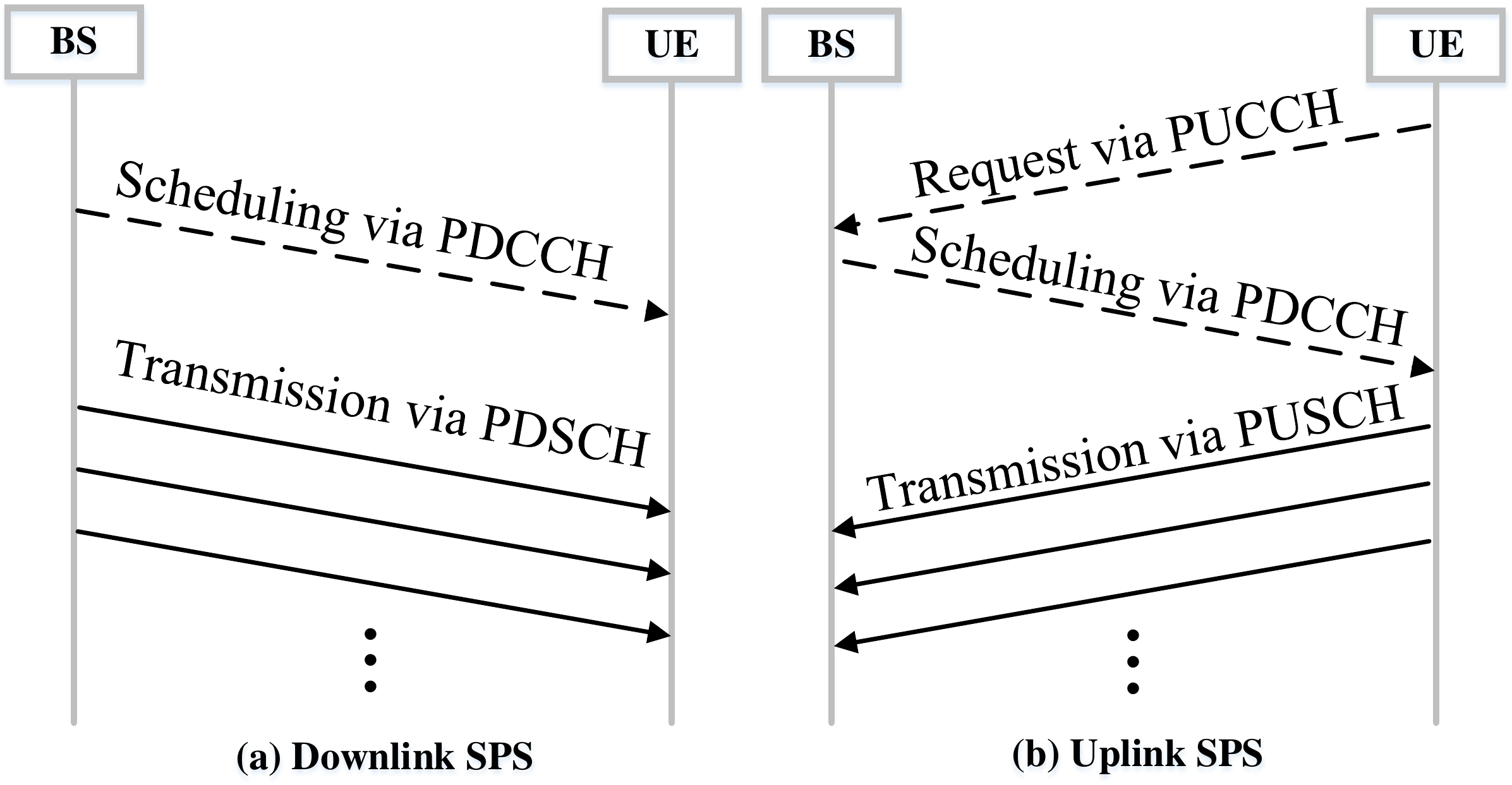}
        \vspace*{-0.4cm}
        \caption{Simplified signaling diagrams of SPS for (a) downlink transmission, and (b) uplink transmission.}
        \label{fig0}
        \end{minipage}
    \end{center}
\vspace*{-0.2cm}
\end{figure}


The original intention of cellular networks is to accommodate bursty traffic generated by human beings, where channel resources, constructed by Orthogonal Frequency Division Multiple Access (OFDMA) frame structure, are mainly allocated using dynamic scheduling (DS) - one packet one allocation process. Apparently, DS is not an efficient scheme to carry small-sized periodic packets, as it creates a significant amount of control signalling overhead and makes large end-to-end latency. To tackle this problem, the SPS technique was developed to serve deterministic traffic by scheduling periodic channel resources in one-shot. As illustrated in Fig. \ref{fig0}, scheduling is triggered by a base station (BS) or request information from an associated user equipment (UE) via Physical Uplink Control CHannel (PUCCH); then, scheduling information is derived by the BS and delivered to the UE via Physical Downlink Control CHannel (PDCCH); finally, periodic packets are transmitted using dedicated channel resources in Physical Downlink/Uplink Sharing CHannel (PDSCH/PUSCH) assigned by the scheduling procedure. The SPS has been employed in several existing cellular systems, e.g., 5G New Radio (NR), Long Term Evolution (LTE), IEEE 802.16e/m, and 3GPP2 UMB, and previously, its main functionality is to support voice over internet protocol (VoIP) services \cite{jiang2007principle}.

Different from human-to-human communication, time-critical wireless applications, especially TSC, raise massive requirements of delivering small-sized periodic data under an ultra-tight delay constraint \cite{IIC2019tsn}. As an example, in motor control applications, periodic two-way traffic is generated by control systems including command signals and positioning feedback, which demands communication services with end-to-end latency in milliseconds level \cite{khoshnevisan20195g}. Apparently, the extremely high scheduling precision from the time domain desired by this traffic exceeds the upper bound of SPS in existing cellular networks. More recently, the urgency of solving the misalignment problem in SPS has been raised in 3GPP report \cite{3GPP190728}. The potential solution can be a single or a mixture of the following techniques:

\begin{itemize}
    \item \textit{Channel overprovisioning}: this method configures extra channel resource for periodic traffic, which requires no specifications changes of the system, but contributes to an extremely inefficient channel utilization.
    
    \item \textit{Multiple SPS configurations}: this method configures multiple SPS parameters for a single traffic flow. However, this method may require plenty of configurations for a single US, thus reduces the number of flows that can be supported by this UE (i.e., the allowed number of configurations is limited).
    
    \item \textit{Periodic shifting}: this method expects to develop a changeable scheduling period varied according to a pre-defined principle to fit any arbitrary period of incoming traffic. This method is likely to largely reduce periodicity misalignment without sacrificing the capacity of flows to be supported simultaneously. However, developing such a shifting principle can be challenging and the solution might be complex.
    
    \item \textit{Efficient retransmission scheme}: this method does not target the misalignment problem itself, but tries to investigate an effective way to execute retransmissions when the misalignment becoming intolerant.
\end{itemize}

Prior studies have mainly investigated SPS enhancements by improving its retransmission efficiency. A pre-scheduled resource allocation technique is proposed in \cite{abreu2017pre}, which provides retransmission opportunities for group UEs without requiring an extra signaling procedure. The group scheduling method is further enhanced in \cite{yuan2020group}, which optimizes the group size and the number of retransmission resources to meet the latency constraints. In the context of TSC, the link adaptation of the SPS scheme is analyzed in \cite{abreu2020scheduling}. Besides SPS, another closely related work is \cite{eisen2020scheduling}, which proposes a time-slicing architecture for the scheduling of industrial control system traffic in 5G networks. More recently, periodic shifting has been raised as the most potential solution for addressing misalignment problem in SPS due to its high flexibility and acceptable complexity \cite{3GPP-R1-1906615,3GPP-R2-1913453}. Accordingly, a periodic shifting scheme has been proposed in \cite{3GPP-R2-1913453}, which shifts the SPS periodicity with a fixed value obtained by studying traffic pattern. However, the performance of this scheme in terms of reducing periodicity misalignment is limited.



\vspace*{-0.0cm}
\section{System Model and Problem Formulation}
\vspace*{-0.0cm}

\begin{figure}
    \begin{center}
    \begin{minipage}[t]{0.47\textwidth}
    \centering
        \includegraphics[width=1\textwidth]{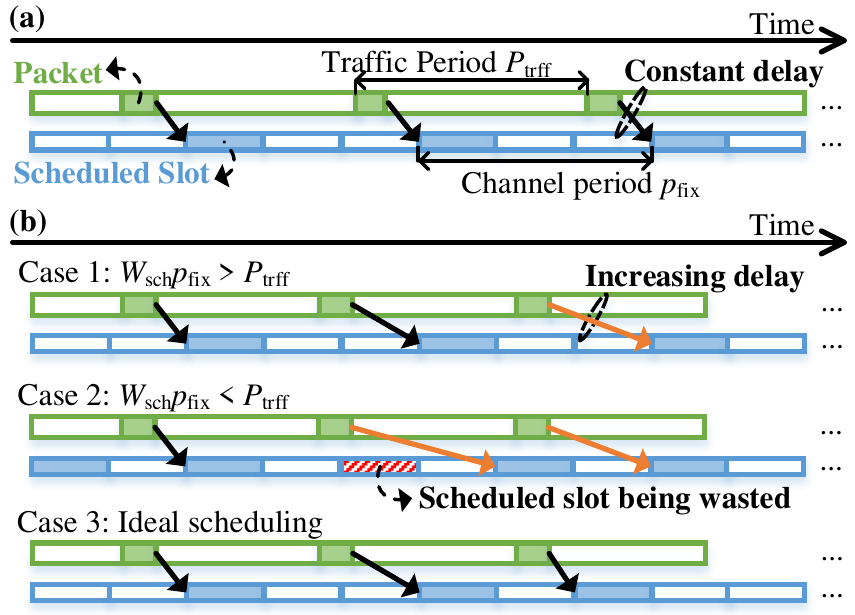}
        \vspace*{-0.2cm}
        \caption{Illustration of matching periodically generated packets to persistently scheduled resource units in (a) period-alignment event, and (b) period-misalignment events.}
        \label{fig1}
        \end{minipage}
    \end{center}
\vspace*{-0.1cm}
\end{figure}

Considering a normalized time scale, the channel scheduling process initializes at a time $\tau_0$, and the periodical packets start to arrive at an arbitrary time $\tau_\text{trff}$ from $[\tau_0, \infty)$. As illustrated in Fig. \ref{fig1}, we consider fixed-cycle traffic, in which packets with the same size are sequentially generated according to a period $P_\text{trff}$. This generation process can occur in both the BS and UE demanding for downlink and uplink transmission, respectively. The transmission is conducted by a network with OFDM-based channel that is divided into slots with equal size $W_\text{sch}$. As we target to tackle the problem of period alignment solely from the temporal domain, the frequency dimension of the channel is omitted. Without loss of generality, we assume the transmission bandwidth is well organized, where the time of two successive transmissions will not be overlapped. Furthermore, to guarantee the stability of the network, we also assume that the traffic period $P_\text{trff}$ is no smaller than the minimal size of a resource unit $W_\text{sch}$.

To serve such periodic traffic, BS performs control signaling before the initial transmission that decides the periodicity of the allocated channel resources. Conventionally, a fixed period $p_\text{fix}$ (measured by slots) are scheduled for serving such periodic traffic. Minimizing misalignment delay and maximizing channel utilization can be achieved at the same time only when the presence of one-to-one correspondence between assignments and incoming packets without unfilled channel. This optimality condition occurs only if the traffic period $P_\text{trff}$ is an integer multiple of the slot length $W_\text{sch}$, namely, \textit{period-alignment} event (as illustrated in Fig. \ref{fig1}(a)). 

However, the numerology options of the network system are limited and unchangeable, which can rarely satisfy that optimality condition. If unsatisfied, the produced interval between two successive assignments $W_\text{sch}p_\text{fix}$ would not equal to the traffic period $P_\text{trff}$, which results in the increase of misalignment delay and wastage of channel resources. In details, denoting an \textit{under-scheduled} state as illustrated in case 1 of Fig. \ref{fig1}(b), the scheduling period $p_\text{fix}$ contributes to the interval $W_\text{sch}p_\text{fix}$ that is slightly bigger than the traffic period $P_\text{trff}$, thus the misalignment delay gradually increases along with time. In opposite, denoting an \textit{over-scheduled} state as shown in case 2 of Fig. \ref{fig1}(b), the scheduling period $p_\text{fix}$ contributes to the interval $W_\text{sch}p_\text{fix}$ that is slightly smaller than the traffic period $P_\text{trff}$. In this case, the assigned slot can be disused due to its appearance in front of the requirement, which leads to a high misalignment delay and the waste of channel resources.

In this work, the objective is to find a close-formed solution, for traffic with arbitrary period value, that can ideally schedule channels in a persistent manner - each assignment appearing right after its corresponding packet (see case 3 in Fig. \ref{fig1}(b)). Accordingly, a changeable period value needs to be adopted for the purpose of postponing or expediting the time of assignments when timing mismatch occurred. To tackle it, we then propose the RPS scheme by solving the following two inter-dependent problems: 1) designing a period varying mechanism, 
for which transceivers can easily derive the exact slot of each assignment; 2) proposing a method for deriving system parameters that can optimally support incoming traffic with arbitrary periodicity.

\vspace*{-0.0cm}
\section{Concept of the RPS Scheme}\label{sec3}
\vspace*{-0.0cm}
In this section, we focus solely on the RPS concept from a mathematical perspective and we leave the problem of how to adopt the RPS concept to a wireless system to Sec. \ref{sec4}. During scheduling, given a traffic period $P_\text{trff}$, the root period $p_\text{0}$ is selected by minimizing the delay that occurs when packets waiting for the assigned slots, which is given as
\begin{align}\label{a-1-1}
p_\text{0} =  \mathop {\text{arg min}}\limits_{ x_0 \in \{1, 2, \cdots, \infty\}}  \mkern9mu \left|P_\text{trff} - x_0W_\text{sch}\right|.
\end{align} 
For arbitrary traffic period, transmission solely relying on the root period can lead to three different states - \textit{period-alignment}, \textit{over-scheduled}, and \textit{under-scheduled}, respectively corresponding to period values $\frac{P_\text{trff}}{W_\text{sch}}$, $\lfloor{\frac{P_\text{trff}}{W_\text{sch}}}\rfloor$ (i.e., $\floor*{\cdot}$ is floor function), and $\lceil{\frac{P_\text{trff}}{W_\text{sch}}}\rceil$ (i.e., $\ceil*{\cdot}$ is ceiling function).


For \textit{over-scheduled} state, the delay that occurs when packets waiting for the assigned slots would no longer be a constant, but continuously decreases with a negative gradient $-\delta_1$, where its value is given by
\vspace*{-0.1cm}
\begin{align}\label{a-1-3}
\delta_1 = |P_\text{trff} - p_0 W_\text{sch}| = P_\text{trff} -\floor{\frac{P_\text{trff}}{W_\text{sch}}} W_\text{sch}. 
\end{align}
Accordingly, one resource unit is wasted every $\frac{W_\text{sch}}{\delta_1}$ assignments. In opposite, for \textit{under-scheduled} state, the delay continuously increases with a positive gradient $\delta_1$, where $W_\text{sch}$ number of delay elements are accumulated every $\frac{W_\text{sch}}{\delta_1}$ assignments. To avoid wasting resources and reduce misalignment delay, we propose a Single-Slot Periodic Alignment (SSPA) mechanism, where, for \textit{over-scheduled} or \textit{under-scheduled} states, the occurrence of all future assignments are periodically postponed or expedited for one single slot every $p_1$ assignments. The alignment period $p_1$ is given as
\begin{align}\label{a-1-4}
p_1 & =\mathop {\text{arg min}}\limits_{ x_1 \in \{1, 2, \cdots, \infty\}}  \mkern9mu \left|W_\text{sch} - x_{1}\delta_{1}\right|.
\end{align}

By aligning the assignments with period $p_1$, the gradient value of the delay reduces to $\delta_2 = |W_\text{sch} - p_{1} \delta_{1}|$. However, if the current gradient value $\delta_2$ was not reduced to $0$, the unsteady delay would still exist. Therefore, given a target of realizing \textit{period-alignment} between assignments and traffic, the scheduling problem is re-formulated as finding a mathematical solution that reduces the gradient value of the delay to $0$. Based on this formulation, in the following, we propose the core strategy of the RPS scheme to mitigate the delay gradient. The strategy is to recursively execute SSPA procedures, where the $n$th ($n>1$) SSPA procedure revises the time of doing the $(n-1)$th SSPA procedure.

\begin{ppro}\label{ppro1}
The alignment period of the $n$th SSPA procedure is given as 
\vspace*{-0.1cm}
\begin{align}\label{a-1-6}
p_n & =\mathop {\text{arg min}}\limits_{ x_n \in \{1, 2, \cdots, \infty\}}  \mkern9mu \left|W_\text{sch} - x_{n}\delta_{n}\right|,
\end{align}
and the instantaneous gradient value of the delay after the $n$th SSPA procedure is obtained by
\begin{align}\label{a-1-7}
\delta_{n+1} =  |p_{n} \delta_{n} - W_\text{sch}|.
\end{align} 
Doing SSPA with Eq. (\ref{a-1-6}) and (\ref{a-1-7}), at least half of the instantaneous delay gradient is cut off in each execution, thus the delay gradient can be gradually reduced to $0$ within the finite number of recursions.
\end{ppro}

\begin{proof}
To minimize the instantaneous gradient of the delay $\delta_{n}$ in the $n$th SSPA procedure, the alignment period $p_n$ can only have two choices that are $\lfloor{\frac{W_\text{sch}}{\delta_{n}}}\rfloor$ and $\lceil{\frac{W_\text{sch}}{\delta_{n}}}\rceil$. In the former case $p_n=\lfloor{\frac{W_\text{sch}}{\delta_{n}}}\rfloor$, the instantaneous gradient of the next recursion given in Eq. (\ref{a-1-7}) can be represented as
\begin{align}\label{pr2}
\delta_{n+1} =  W_\text{sch} - \floor*{\frac{W_\text{sch}}{\delta_{n}}} \delta_{n}
\le  W_\text{sch} - \frac{W_\text{sch}-\frac{\delta_{n}}{2}}{\delta_{n}} \delta_{n}
=\frac{\delta_{n}}{2}.
\end{align} 
Similarly, if $p_n=\lceil{\frac{W_\text{sch}}{\delta_{n}}}\rceil$, the instantaneous gradient of the next recursion given in Eq. (\ref{a-1-7}) can be represented as
\begin{align}\label{pr3}
\delta_{n+1} =  \ceil*{\frac{W_\text{sch}}{\delta_{n}}} \delta_{n}-W_\text{sch}
\le  \frac{W_\text{sch}+\frac{\delta_{n}}{2}}{\delta_{n}} \delta_{n} - W_\text{sch}
=\frac{\delta_{n}}{2}.
\end{align} 
Accordingly, we find the instantaneous gradient value of the delay in any scenarios following $\delta_{n+1}\le\frac{\delta_{n}}{2}$. Therefore, the RPS scheme can always find a solution that reduces the gradient to $0$ within a finite number of recursions.
\end{proof}



\section{RPS Scheme Design in Cellular Networks}\label{sec4}
In this section, we propose an implementation-friendly method to apply the RPS scheme in cellular networks. Note that, for the novel RPS scheme, the scheduling and transmission procedures are still based on the conventional way as shown in Fig. \ref{fig0}, only the derivation of channel resource periodicity is changed. We first propose a period alignment mechanism in Sec. \ref{sec4.1} that can enable RPS concept in a cellular environment, where we answer the following two questions: 1) what type of scheduling parameters are required, and 2) How to exactly derive positions of assigned resources based on the known scheduling parameters during transmission. According to this cellular-enabled RPS mechanism, we then, in Sec. \ref{sec4.2} and Sec. \ref{sec4.3}, propose methods for deriving scheduling parameters with any resource window size $W_\text{sch}$ under arbitrary traffic period $P_\text{trff}$ that can perfectly accommodate traffic with minimal periodicity misalignment.

\subsection{Scheduling and Transmission Mechanism of RPS Scheme }\label{sec4.1}

To serve periodic traffic, the BS needs to calculate a tuple of scheduling factors ${\cal A}_N=\{{{\vec \alpha}_0,{\vec\alpha}_1,{\vec\alpha}_2,⋯,{\vec\alpha}_N}\}$ according to the prior information of traffic pattern, including the traffic initial time $\tau_\text{trff}$ and traffic period $P_\text{trff}$. To do so, the BS first derives a set of root factors ${\vec \alpha}_\text{0} = \{p_0, q_0, t_0\}$, referring to the scheduling period of resource units (a.k.a. root period), the alignment direction factor of the root period, and the initial position of assignments, respectively. To tackle the problem of periods-mismatching, the BS then derives a timing-alignment policy, which revises the root period $p_\text{0}$ by postponing or expediting the scheduled resource units. This timing-alignment policy consists of $N$ number of SSPA procedure each with a set of factors ${\vec \alpha}_n = \{p_n, q_n, t_n\}$ w.r.t. the alignment period, the direction factor of each alignment step, and the exact slot that triggers the alignment procedure (a.k.a., initial position). The recursive process is ceased after $N$ alignments on meeting the optimality condition - the $N$th instantaneous gradient value of the delay $\delta_{N}=0$.


During scheduling, transceivers are informed by scheduling factors ${\cal A}_N$. Based on these factors, the exact slots of assignments are derived by transceivers to accommodate the transmission/reception of the relevant packets. The slot index of the $m$th assignment is obtained by
\begin{align}\label{a-2-2}
\varphi^m_\text{umt} = t_\text{0} + (m-1) p_\text{0} +  q_0 \phi^m_1,
\end{align}
where $m \in \{1, 2, \cdots, M \}$, and $M$ is the total number of packets required to be served. Note that the scheduling policy in the period-alignment case can be easily obtained by setting $q_0 \phi^m_1=0$. In Eq. \eqref{a-2-2}, a binary variable $q_0\in\{-1, 1\}$ is used to decide the alignment direction of the root period $p_0$. And, $\phi^m_1$ is the exact number of slots to be revised for the $m$th assignment, which is obtained via a finite recursion process. This is given as
\begin{align}\label{a-2-3}
\phi^m_n = \floor*{ \frac{p_n - t_n + m - q_{n}\phi^m_{n+1}}{p_1} }.
\end{align}
The recursion process in (\ref{a-2-3}) is calculated according to the order $\{N-1,N-2,\cdots, 1\}$, and is initialized by inputting $\phi^m_N$ that is given by
\begin{align}\label{a-2-4}
\phi^m_N =  \floor*{ \frac{p_N - t_N + m }{p_N} }.
\end{align}

The procedure that derives the position of assigned resource for the $m$th packet during transmission/reception in the RPS scheme is shown in \textbf{Algorithm \ref{a1}}. 

\begin{algorithm}[t]
\footnotesize
\SetKwData{Left}{left}
\SetKwData{This}{this}
\SetKwData{Up}{up}
\SetKwFunction{Union}{Union}
\SetKwFunction{FindCompress}{FindCompress}
\SetKwInOut{Input}{input}
\SetKwInOut{Output}{output}
\caption{Deriving slot index of  $m$th assignment} \label{a1}
\Input{Scheduling factors ${\cal A}_N=\{{{\vec\alpha}_0,{\vec\alpha}_1,{\vec\alpha}_2,⋯,{\vec\alpha}_N}\}$.}
Calculate $\phi^m_N$ using Eq. (\ref{a-2-4}) with $\vec{\alpha}_N$; \\
\For{$n \leftarrow$ ($N-1$) \KwTo $1$}{
Calculate $\phi^m_n$ using Eq. (\ref{a-2-3}) with $\phi^m_{n+1}$ and $\vec{\alpha}_n$; \\
}
Calculate $\varphi^m_\text{umt}$ using Eq. (\ref{a-2-2}) with $\phi^m_{1}$ and $\vec{\alpha}_0$;\\
\end{algorithm}

\subsection{Derivation of Alignment Periods and Direction Factors}\label{sec4.2}

During scheduling, we first calculate the root period $p_\text{0}$ using Eq. (\ref{a-1-1}) according to the given traffic period $P_\text{trff}$. If the condition $[P_\text{trff} \mkern9mu  \text{mod} \mkern9mu W_\text{sch}] = 0$ was satisfied, the period alignment would not be activated. Otherwise, the RPS mechanism requires to be activated, and the alignment direction factor of the root period is given as
\begin{align}\label{a-3-4}
q_0 =  \left\{
\begin{aligned}
& 1, \mkern27mu  & & 0<[P_\text{trff} \mkern9mu \text{mod} \mkern9mu W_\text{sch}] \le \floor*{\frac{W_\text{sch}}{2}},
\\
& -1,  \mkern27mu & &[P_\text{trff} \mkern9mu \text{mod} \mkern9mu W_\text{sch}]> \floor*{\frac{W_\text{sch}}{2}}.
\end{aligned}\right.
\end{align}
In (\ref{a-3-4}), $q_0=1$ and $q_0=-1$ respectively refers to the activity of postponement and
expedition.

The recursive alignment procedure is terminated if the step 1 alignment can mitigate the timing mismatch between the incoming traffic and assigned channels, otherwise, an $N$-steps recursive alignment requires to be executed. Note that, different from the step 1 alignment, the alignment in each step $n$ ($n>1$) revises the time of doing the step $n-1$ alignment instead of the time of assignment as elaborated in Sec. \ref{sec3}. According to \textbf{Proposition \ref{ppro1}}, the period of alignment for any step $n$ is given as
\begin{align}\label{a-3-5}
p_n & = \left\{
\begin{aligned}
&\frac{W_\text{sch}}{\delta_{n}},  \mkern27mu & &[W_\text{sch} \mkern9mu  \text{mod} \mkern9mu \delta_{n}] = 0 , 
\\
&\floor*{\frac{W_\text{sch}}{\delta_{n}}},  \mkern27mu  & & 0<[W_\text{sch} \mkern9mu \text{mod} \mkern9mu \delta_{n}] \le \floor*{\frac{\delta_{n}}{2}},  
\\
&\ceil*{\frac{W_\text{sch}}{\delta_{n}}},   \mkern27mu & &[W_\text{sch} \mkern9mu \text{mod} \mkern9mu \delta_{n}]> \floor*{\frac{\delta_{n}}{2}}.
\end{aligned}\right.
\end{align}
Occasionally, one may meet a scenario that the selection of either $\lfloor{\frac{W_\text{sch}}{\delta_{n}}}\rfloor$ or $\lceil{\frac{W_\text{sch}}{\delta_{n}}}\rceil$ experiencing the same performance. For clarity, we only consider the bigger selection in the proposed method. In Eq. (\ref{a-3-5}), the instantaneous gradient of the delay $\delta_n$ at step $n$ alignment is given by Eq. (\ref{a-1-7}) in \textbf{Proposition \ref{ppro1}}. And the direction factor $q_n$ of the $n$th SSPA procedure is obtained by 
\begin{align}\label{a-3-7}
q_n =  \left\{
\begin{aligned}
& 1, \mkern16mu  \mkern27mu  & & 0<[W_\text{sch} \mkern9mu \text{mod} \mkern9mu \delta_{n}] \le \floor*{\frac{\delta_{n}}{2}},  
\\
& -1, \mkern9mu  \mkern27mu & &[W_\text{sch} \mkern9mu \text{mod} \mkern9mu \delta_{n}]> \floor*{\frac{\delta_{n}}{2}}.
\end{aligned}\right.
\end{align}
The procedure of recursive alignments is terminated when satisfying the condition $[W_\text{sch}$ mod $\delta_{n}]$ $= 0$. The derivation process of alignment periods and direction factors during scheduling in the RPS scheme is detailed in \textbf{Algorithm \ref{a2}}.

\begin{algorithm}[t]
\footnotesize
\SetKwData{Left}{left}
\SetKwData{This}{this}
\SetKwData{Up}{up}
\SetKwFunction{Union}{Union}
\SetKwFunction{FindCompress}{FindCompress}
\SetKwInOut{Input}{input}
\SetKwInOut{Output}{output}
\caption{Deriving alignment periods and direction factors} \label{a2}
\Input{Traffic period $P_\text{trff}$.}
Calculate the root period $p_0$ using Eq. (\ref{a-1-1});\\
\If{$[P_\text{trff} \mkern9mu  \text{mod} \mkern9mu W_\text{sch}] \neq 0 $}
{
Calculate the direction factor $q_0$ using Eq. \eqref{a-3-4}; \\
Set $n=1$; \\
\Repeat( $n=n+1$)
{$[W_\text{sch} \mkern9mu  \text{mod} \mkern9mu \delta_{n}] =0$}
{
Calculate the gradient value $\delta_{n}$ using Eq. \eqref{a-1-3}$/$\eqref{a-1-7}; \\
Calculate the alignment period $p_{n}$ using Eq. \eqref{a-3-5}; \\
Calculate the direction factor $q_n$ using Eq. \eqref{a-3-7}; \\
}
Obtain the required number of alignments $N=n$;
}
\end{algorithm}

\subsection{Derivation of Initial Positions}\label{sec4.3}

The initial position of root scheduling $t_\text{0}$ can be easily known by counting the number of transmission windows within duration $[\tau_0, \tau_\text{trff}]$, which is given by
\begin{align}\label{a-3-8}
t_\text{0} =  \ceil*{\frac{\tau_\text{trff}-\tau_0}{W_\text{sch}}} + 1.
\end{align}

For each alignment procedure ($n=1,2,\cdots,N$), the initial position $t_n$ is decided according to the prior initial position $t_{n-1}$, which has $p_n$ number of potential choices. To obtain the optimal combination of every initial position, one can evaluate all possibilities by using exhaustive search algorithm. However, given the number of alignment steps $N$ that can be large, the exhaustive search procedure would be resource-hungry. To tackle this problem, we propose a heuristic algorithm, namely, Joint Derivation and Verification (JDV) algorithm, which greatly simplifies the derivation procedure of initial position $t_n$.

The JDV algorithm is given in \textbf{Algorithm \ref{a3}}, in which two successive processes - \textit{forward derivation} and \textit{backward validation}, are developed for the derivation of each initial position $t_n$. We denote ${\cal A}_n=\{{{\vec\alpha}_0,{\vec\alpha}_1,{\vec\alpha}_2,⋯,{\vec\alpha}_n}\}$ that is a subset of the target tuple ${\cal A}_N$, which is composed of the first $n+1$ vectors in ${\cal A}_N$. For the derivation of any $n$th value, we first calculate an estimated initial value $\tilde{t}_n$ by using the known parameters in ${\cal A}_{n-1}$ in the \textit{forward derivation} process. To do so, we calculate the gap between $\varphi_\text{umt}^m (m=i|{{\cal A}}_{n-1})$ (obtained via Eq. (\ref{a-2-2})) and the exact arriving time of the packet at any slot $i$, under the condition of input parameters with ${\cal A} = {{\cal A}}_{n-1}$, which is given as 
\vspace*{-0.2cm}
\begin{align}\label{a-3-9}
\theta_n (m|{\cal A}) = & (\varphi_\text{umt}^m (m|{\cal A})-1) W_\text{sch} 
- (\tau_\text{trff}-\tau_0) - (m-1)P_\text{trff}.
\end{align}
We obtain the estimated initial value $\tilde{t}_n=i$ until satisfying the condition of $\theta_n (i|{\cal A}_{n-1})<0$ or $\theta_n (i|{\cal A}_{n-1})≥W_\text{sch}$.


\begin{algorithm}[t]
\footnotesize
\SetKwData{Left}{left}
\SetKwData{This}{this}
\SetKwData{Up}{up}
\SetKwFunction{Union}{Union}
\SetKwFunction{FindCompress}{FindCompress}
\SetKwInOut{Input}{input}
\SetKwInOut{Output}{output}
\caption{JDV-Deriving initial position $t_n$} \label{a3}
\Input{Traffic period $P_\text{trff}$, alignment periods $\{p_0,p_1,\cdots,p_n\}$.}
Calculate initial position of the root scheduling $t_0$ using Eq. (\ref{a-3-8}); \\
\For{$n \leftarrow 1$ \KwTo $N$}{
$i = 2$ \textbf{if} $n=1$ \textbf{else} $t_{n-1}$;\\
\textit{$/*$ Forward derivation: \hfill $*/$}\\
\Repeat( $i=i+1$)
{$\theta_n (i|{\cal A}_{n-1})<0$ \textbf{or} $\theta_n (i|{\cal A}_{n-1})≥W_\text{sch}$}
{
Calculate $\varphi^m_\text{umt}(m=i|{\cal A}_{n-1})$ using Eq. (\ref{a-2-2});\\
Calculate $\theta_n (i|{{\cal A}}_{n-1})$ using Eq. (\ref{a-3-9});\\
}
\textit{$/*$ Backward validation: \hfill $*/$}\\
Set $\tilde{t}_n = i$;\\
\Repeat( $\tilde{t}_n=\tilde{t}_n-1$)
{$0\le\theta_n (j|\widetilde{{\cal A}}_n)<W_\text{sch}$, $\forall j\in[\tilde{t}_n,\tilde{t}_n+\gamma p_n+1]$ }
{
\If{${t}_{n-1} >\tilde{t}_n$}{Set ${t}_{n-1}=\tilde{t}_n$;}
Set $\widetilde{{\cal A}}_n = \{{\vec\alpha}_0,{\vec\alpha}_1,{\vec\alpha}_2,⋯,{\vec\alpha}_{n-1}, {\tilde \alpha}_n \}$ with $\tilde{t}_n$;\\
\For{$j \leftarrow \tilde{t}_n$ \KwTo $\tilde{t}_n+\gamma p_n+1$}{
Calculate $\varphi^m_\text{umt}(m=j|\widetilde{{\cal A}}_n)$ using Eq. (\ref{a-2-2});\\
Calculate $\theta_n (j|\widetilde{{\cal A}}_n)$ using Eq. (\ref{a-3-9});\\
}
}
Obtain $t_n = \tilde{t}_n$;\\
}
\end{algorithm}

We then develop a \textit{forward derivation} process to validate the correctness of $\tilde{t}_n$, and corrects it if the mismatch occurs between the assignment and traffic. To do so, we first construct a temporary tuple $\widetilde{{\cal A}}_n$ including the exact factors $\{{{\vec\alpha}_0,{\vec\alpha}_1,{\vec\alpha}_2,⋯,{\vec\alpha}_{n-1}}\}$ and the vector ${\tilde \alpha}_n = \{p_n, q_n, \tilde{t}_n\}$ with the estimated initial position $\tilde{t}_n$ obtained in the \textit{forward derivation} process. Then, we calculate the gap between $\varphi_\text{umt}^m (m=j|\widetilde{{\cal A}}_{n})$ (obtained via Eq. (\ref{a-2-2})) and the exact arriving traffic, under the condition of input parameters with ${\cal A} = \widetilde{{\cal A}}_{n}$ via Eq. (\ref{a-3-9}). Finally, we evaluate if gap $\varphi_\text{umt}^m (m=j|\widetilde{{\cal A}}_{n})$ satisfies the condition $0\le\varphi_\text{umt}^m (m=j|\widetilde{{\cal A}}_{n})<W_\text{sch}$ for any $j\in [\tilde{t}_n, \tilde{t}_n+\gamma p_n + 1]$, otherwise, we set $\tilde{t}_n=\tilde{t}_n-1$, until obtaining a satisfied initial position. Note that, the factor $\gamma$ in the range $[\tilde{t}_n, \tilde{t}_n+\gamma p_n + 1]$ is heuristically selected. According to our simulation, selecting $\gamma=3$ can guarantee the ideal assignment for any traffic scenario when $W_\text{sch}<1000$ $\mu$s, which satisfies the requirements of the most cellular networks. 


\vspace*{-0.0cm}
\section{Performance Evaluation}

We have conducted numerical simulations to evaluate the misalignment delay of the RPS-SPS scheme. We consider a network with the channel slotted into fixed-duration units of $W_\text{sch}=0.071$ ms (i.e., duration of 2 OFDM symbols with 30 kHz SCS in 5G NR \cite{elayoubi2019radio}). 
For comparison, two baseline schemes are considered, namely, PS-SPS and C-SPS, referring to fixed-value periodic shifting SPS proposed in \cite{3GPP-R2-1913453} and classical SPS in existing cellular systems \cite{jiang2007principle}. During simulations, a packet is dropped when it experiencing a misalignment delay that is longer than its period length.

\begin{figure}
    \begin{center}
    \begin{minipage}[h]{0.49\textwidth}
    \centering
        \includegraphics[width=1\textwidth]{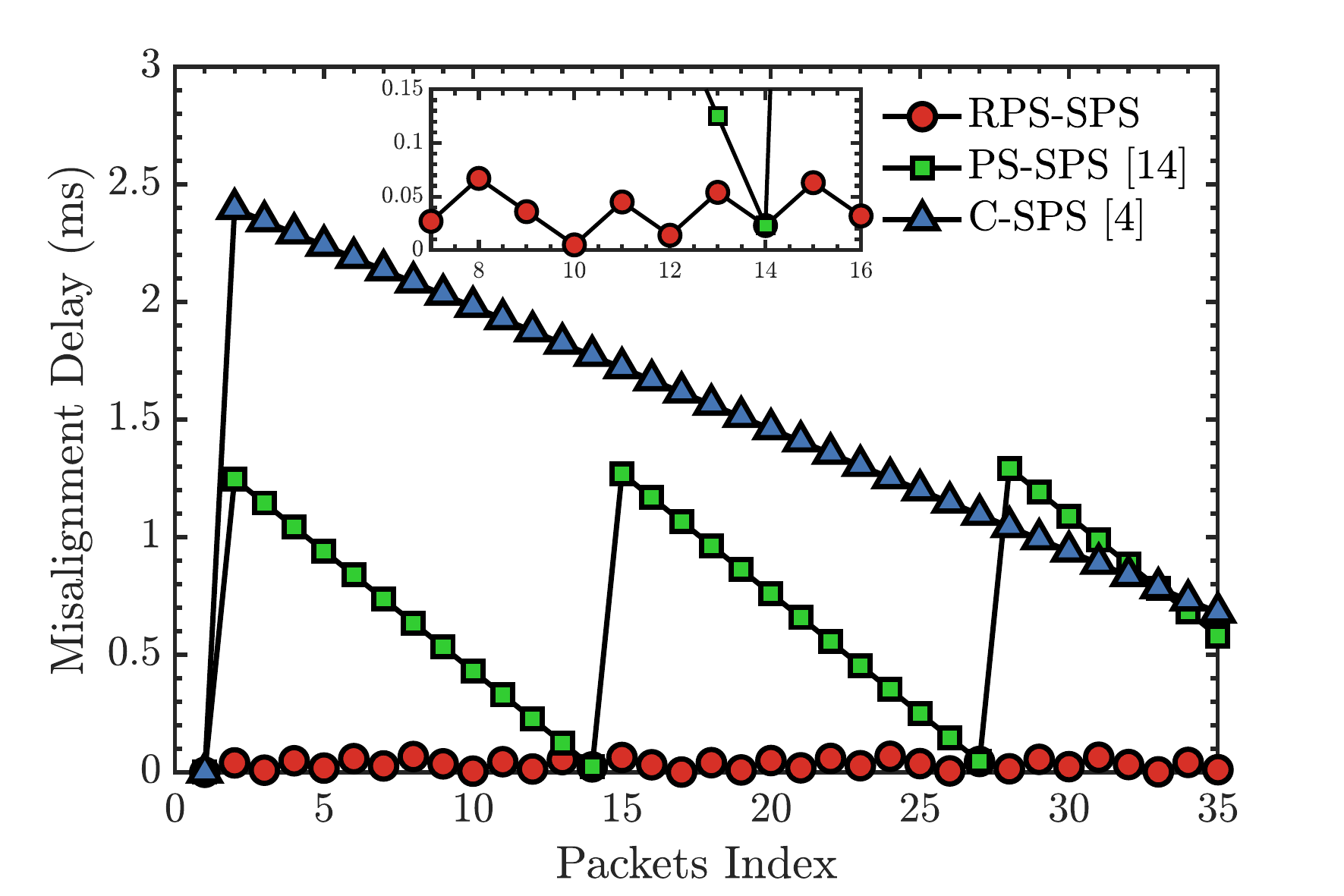}
      \vspace*{-0.5cm}
        \caption{Simultaneous misalignment delay versus the index of incoming packets during one traffic duration.}
                \label{fig3}
        \end{minipage}
    \end{center}
\vspace*{-0.2cm}
\end{figure}

We start by demonstrating the operation of the proposed and reference scheduling schemes in Fig. \ref{fig3}. This figure plots the simultaneous misalignment delay of each transmission along the order of incoming packets, which are periodically generated according to traffic period $P_\text{trff}=2.8$ ms. This period generally captures patterns of critical isochronous traffic in TSC applications, e.g., motion control \cite{khoshnevisan20195g}. We observe that the misalignment delay of the proposed RPS scheme is always smaller than the duration of resource unit (0.071 ms). This is due to the proposed RPS-SPS to perfectly schedule resource units that each assignment locates right after its corresponding packet arriving. In opposite, the misalignment delay of reference schemes are generally higher than the duration of resource unit. We further observe two interesting trends of the reference performance - the delay first continuously decreases with a fixed gradient until reach the bottom, and then sharply increases to the peak. The former trend appears as the period of assignments is smaller than that of the traffic, while the latter trend appears as one assigned slot is disused due to its appearance in front of the packet incoming.


\begin{figure}
    \begin{center}
    \begin{minipage}[h]{0.49\textwidth}
    \centering
        \includegraphics[width=1\textwidth]{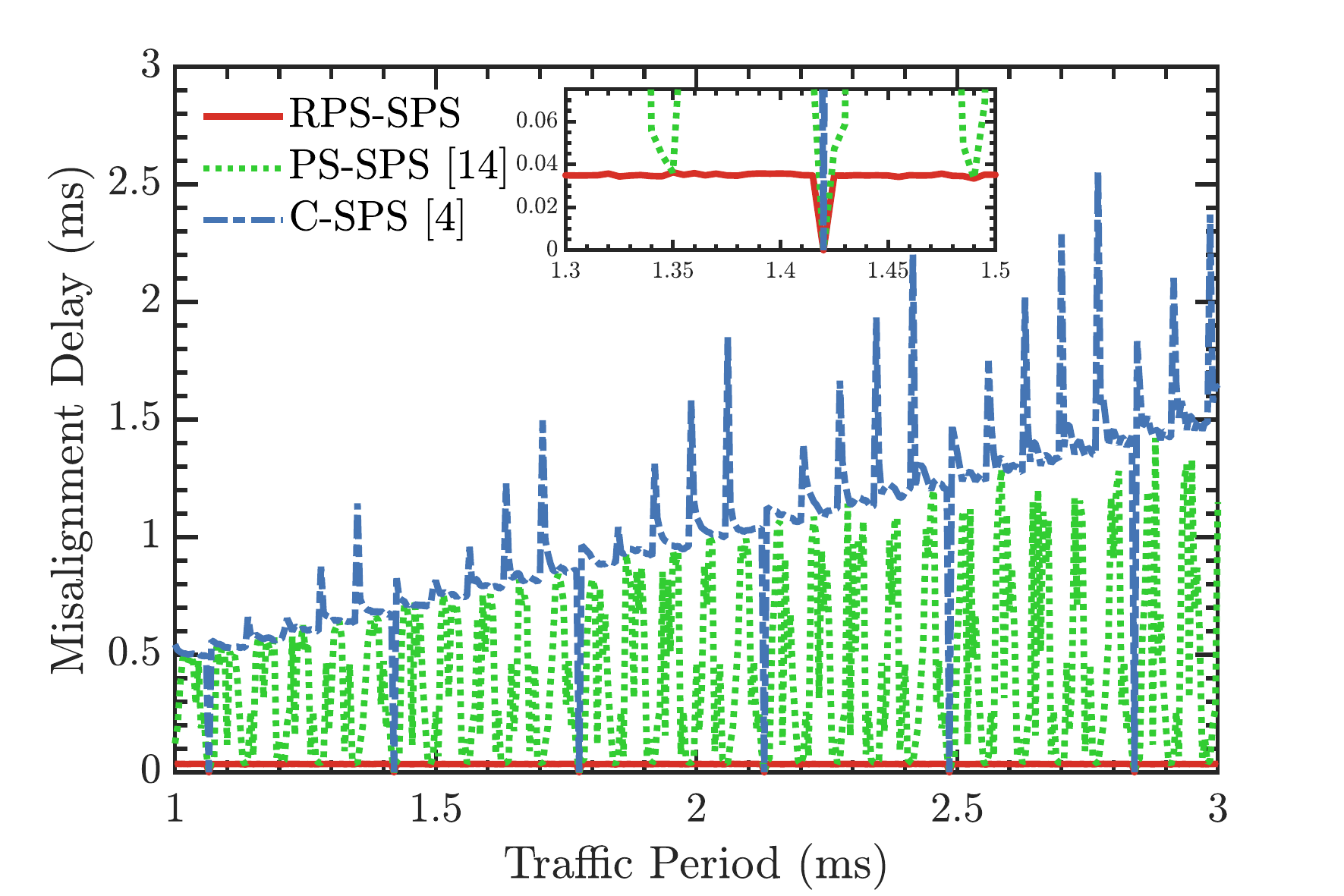}
        \vspace*{-0.5cm}
        \caption{Average misalignment delay per packet versus the packet generation period of deterministic traffic.}
                \label{fig4}
        \end{minipage}
    \end{center}
\vspace*{-0.2cm}
\end{figure}

Figure \ref{fig4} plots the average misalignment delay per packet over 200 transmissions versus the traffic period $P_\text{trff}$ of 1 to 3 ms with increment of 5 $\mu$s. We observe that increasing traffic period $P_\text{trff}$ aggravates the misalignment delay of reference schemes, since a larger traffic period leads to a lower density of assignments, which increases the upper bound of the instantaneous misalignment delay. Conversely, the misalignment delay of RPS-SPS is not affected by the increase of traffic period, which always maintain at a level smaller than the duration of resource units (0.071 ms). 
This demonstrates the capability of the RPS-SPS scheme to exactly match the time between assignments and incoming packets for any periodic traffic statistics.

\vspace*{-0.3cm}
\section{Conclusion}

We have developed a novel RPS scheme for time-sensitive communication in 5G/6G that minimizes the misalignment delay arising from timing mismatch between the allocated resource units and incoming periodic data packets. In the proposed scheme, an SSPA mechanism was developed to enable the variability of the scheduling period, which is recursively executed to find the optimal scheduling policy that can accurately assign resource units located right after their corresponding packets. We mathematically prove the convergence of the RPS scheme, and proposed an implementation-friendly method that adopts the RPS scheme to existing cellular systems. Performance evaluation demonstrate that the proposed scheme can always achieve the minimal misalignment delay with any periodic traffic statistics.


\vspace*{-0.0cm}
\bibliographystyle{IEEEtran}
\bibliography{IEEEabrv,TSN_bib}

\end{document}